\documentstyle [12pt]{article}
\makeatletter \oddsidemargin  0in \evensidemargin 0in
\textwidth16cm \RequirePackage[dvips]{graphicx} \textheight 20.5cm
\newcommand{\singlespacing}{\let\CS=\@currsize\renewcommand{\baselinestretch}{1.5}\tiny\CS}
\makeatletter \oddsidemargin  -.1in \evensidemargin -.1in
\newcommand{\doublespacing}{\let\CS=\@currsize\renewcommand{\baselinestretch}{1.35}\tiny\CS}

\def\@citex[#1]#2{\if@filesw\immediate\write\@auxout{\string\citation{#2}}\fi
  \def\@citea{}\@cite{\@for\@citeb:=#2\do
    {\@citea\def\@citea{,\linebreak[0]\hskip0pt plus .2em}%
      \@ifundefined{b@\@citeb}%
    {{\bf ?}\@warning{Citation `\@citeb' on page \thepage\space undefined}}%
      \hbox{\csname b@\@citeb\endcsname}}}{#1}}

\newtheorem{rule-def}[theorem]{Rule}
\begin{document}
\title{\bf Two Party Non-Local Games}\author{I.Chakrabarty $^{1,2}$\thanks{Corresponding author:
E-Mail-indranilc@indiainfo.com }, B.S.Choudhury $^2$, \\
$^1$ Heritage Institute of Technology,Kolkata-107,West Bengal,India\\
$^2$ Bengal Engineering and Science University, Howrah, West
Bengal, India }
\date{}
\maketitle{}
\begin{abstract}In this work we have introduced two party
games with respective winning conditions. One cannot win these
games deterministically in the classical world if they are not
allowed to communicate at any stage of the game. Interestingly we
find out that in quantum world, these winning conditions can be
achieved if the players share an entangled state. We also
introduced a game which is impossible to win if the players are
not allowed to communicate in classical world (both
probabilistically and deterministically), yet there exists a
perfect quantum strategy by following which, one can attain the
winning condition of the game.
\end{abstract}
\section{Introduction:} It is well known that in quantum
information theory the amount of communication required to perform
various distributed computational tasks is much less than its
classical protocol [1,2]. Given a prior share of entanglement
between two spatially separated parties, the amount of classical
communication required for distributed computational tasks can be
reduced to a large extent. Inspired by the nonlocal properties of
the entangled states various pseudo telepathy games were
introduced. The study of pseudo telepathy games was first
introduced in [2], although the term pseudo-telepathy was first
introduced in ref [6]. Later a multi party pseudo-telepathy game
was introduced [4]. The first explicit quantum pseudo telepathy
game was proposed by Vaidman [3] as a minor variation of Ref. [5].
Indeed, Vaidman was also the first to suggest [11] the first
two-player quantum pseudo telepathy game as a variation of
Cabello's proof [12]. The well known Pseudo-telepathy games are
'Impossible Colouring Game'[5,6], 'Parity Game' [4,6],
'Deutsch-Jozsa Game'[3,6], 'Matching Game' [7], 'Magic Square
game'[8,9] and so many. A different two-player quantum pseudo
telepathy game has been recently introduced in [13].
\\
In order to formulate a multi party pseudo telepathy game let us
consider n parties $X_1,X_2,X_3,....,X_n$ and two n-ary functions
$f$ and $g$. Initially they are allowed to discuss their strategy
and game plan (in classical situation)and also entanglement for
quantum case. Once they are separated, they are strictly
prohibited from any further communication. After that they are
provided with inputs $a_i$ and asked to produce the output $b_i$.
The players are said to be winners if they satisfy the winning
condition $f(a_1,a_2,....,a_n)=g(b_1,b_2,....,b_n)$. For a
particular game we define a n ary predicate $P$ known as the
promise. We say that the protocol is perfect if the promise is
satisfied. There are several reasons for which pseudo-telepathy
games are appealing to us.\\
$\textbf{1.}$ Conceptually these games are very simple and
comprehensible.\\
$\textbf{2.}$ If these games are carried out successfully, these
will provide strong evidence of the fact that quantum physics can
be harnessed to reduce the amount of communication required to
perform various distributive works.\\
However there are certain features for which the experimental
demonstration of pseudo telepathy games are not appealing to the
physicists.\\
$\textbf{1.}$ A small margin of error in the implementation of the
game would result in very high error probability than the
corresponding classical protocol.\\
$\textbf{2.}$ The evidence of convincing quantum behavior would
require a large number consecutive successful runs.\\
In this work we introduced few two party non local games . These
games are different from pseudo-telepathy games in the sense that,
these games unlike pseudo-telepathy games do not allow any kind of
communication between the players before the onset of the games.
It is a well known fact that pseudo telepathy games do not exist
for bi-partite entangled states [10]. Classical players, without
communicating with one another, cannot achieve the winning
condition of these games deterministically, where as the quantum
players with prior share of the entanglement can exactly do so. In
this work we claim to develop a two party game which is impossible
to win in the classical world (both deterministically and
probabilistically), and interestingly we find a quantum strategy
by which this game can be won with a prior share of entanglement
between these two players.
\section{Two party Non-Local Games: Perfect Quantum strategies}
{\bf Game 1:} Let us consider two players Alice and Bob who are
separated from each other in such a way so that they cannot
communicate with each other. Before these two players are
separated they are not allowed to communicate between each other
so that they can share random variables and also not allowed to
discuss among themselves to determine a particular strategy to
win the game. Once they are separated far apart they are devoid
of any kind of communication between each other. At a certain
instant the two spatially separated parties are asked a question
$\textbf{X}$, which has two possible answers $\{ \textbf{+1,-1}
\}$. Interestingly the same question will be asked to both
players for n times. Not only that each of the players are
deprived of the memory of the answers given by them in previous
trials. This game has no such promise which has to be fulfilled
by the inputs or in other words no such restrictions are given on
the legitimacy of the question asked to both the players . The
players are said to win the game if the sum of all the outputs
given by both the players at the end of all the n possible trials
adds to zero. Mathematically, the winning condition is given by,
\begin{eqnarray}
\sum_i^n (X_{i}^{A}+X_{i}^{B})=0
\end{eqnarray}
where $X_{i}^{A}$ and $X_{i}^{B}$ are the answers provided by
Alice and Bob at the ith trial respectively. In classical setting
the game cannot be won deterministically. However the winning
condition can
be achieved by the classical players at least probabilistically.\\
If $n=1$, the total number of outcomes are 4 and the probability
of obtaining the sum $0$ is $\frac{1}{2}$. All possible outcomes
are given by,
\begin{eqnarray}
+1-1=0\nonumber\\
+1+1=2\nonumber\\
-1-1=2\nonumber\\
-1+1=0.
\end{eqnarray}
If $n=2$ all possible outcomes of the game are 16. The number of
cases favorable to the event of obtaining the sum of the answers
to be $0$ are,
\begin{eqnarray}
+1+1-1-1=0\nonumber\\
+1-1+1-1=0\nonumber\\
-1+1+1-1=0\nonumber\\
+1-1-1+1=0\nonumber\\
-1+1-1+1=0\nonumber\\
-1-1+1+1=0.
\end{eqnarray}
It is clearly evident from equation (3) the probability for
winning the game classically is $\frac{6}{16}$. A little
mathematical calculation will show that for n trials, the total
number of outcomes are $2 ^{2n}$. The number of cases which are
favorable to the event of winning the game are $^{2n}C_{n}$. Hence
the probabibility of obtaining the sum answers given by two
parties after n trials as $0$ is $\frac{^{2n}C_{n}}{2 ^{2n}}$. As
$n$ tends to a large number the probability of the event tends to
$0$, i.e $\lim_{n\rightarrow \infty} \frac{^{2n}C_{n}}{2
^{2n}}=0$. So it is clearly evident that the game described above
cannot be done in
the classical world deterministically.\\
However there exists a perfect quantum solution to the game.
Initially Alice and Bob are allowed to share an entangled state
and decide the quantum strategy so that their outputs can attain
the winning condition of the game. Let us assume that Alice and
Bob share an entangled state,
\begin{eqnarray}
|\psi\rangle= \frac{1}{\sqrt{2}}[|01\rangle-|10\rangle]
\end{eqnarray}
(where the first qubit belongs to Alice whereas the second qubit
belongs to Bob). The strategy to win the game is like this : If a
member is asked the $\textbf{X}$ question, he/she will measure the
spin of his/her qubit along x direction, i.e he/she will measure
$\sigma_{x}$.Measurement results (+1 and -1 eigenvalues) will be
the answers. Now it is clearly evident that the state
$|\psi\rangle$ has an eigen value $0$ for the observable
$\sum_{i=1}^{n} (\sigma_{x,i}^A+\sigma_{x,i}^B)$ (where $i$
denotes the number of trials),
\begin{eqnarray}
\sum_{i=1}^{n}
(\sigma_{x,i}^A+\sigma_{x,i}^B)|\psi\rangle=0|\psi\rangle
\end{eqnarray}
Hence we can conclude that they can win the game in every possible
cases. Thus we see that for above described game it is impossible
for the classical players to win the game deterministically,
whereas there exists a quantum strategy for quantum players to
achieve the winning condition.\\
{\bf Game 2:} In this subsection we will consider another
pseudo-telepathy game, impossible to win deterministically in the
classical world, but still admits of a quantum solution. Let us
consider two friends Alice and Bob who are not initially allowed
to interact so that they can share whatever information they like
to. Even when  they are separated , they are not allowed to
communicate any kind of information between them. Once again each
of them are asked a question $X$ for $n$ number of times. The
solutions to the question are $\{+1,-1\}$. Each of the players
answers in any particular trial doesn't depend on what they have
answered in previous trials. The game has been so designed that
the players are declared as the winners only when the product of
the outputs of the two players at the end of n trials is equal to
1. Mathematically, the winning condition is given by,
\begin{eqnarray}
\prod_{i=1}^{n}X_i^AX_i^B=1
\end{eqnarray}
where $X_{i}^{A}$ and $X_{i}^{B}$ are the outcomes of Alice and
Bob at the ith trial respectively. Classically the distant players
do not have a deterministic solution to the above game. But still
one can win the game with certain probabilities of success. A
straight forward observation will reveal that the probability
obtaining the product of the outcomes by two players at the end of
n trials to be 1 is $\frac{1}{2}$. Whatever be the number of
trials , the probability will not decrease like the previous game
and will remain as a constant quantity $\frac{1}{2}$.\\
However there exists a quantum strategy under which the winning
condition can be achieved by the players deterministically if they
share an entangled state initially. The quantum players are not
allowed to communicate with each other at any stage of the game.
Let us consider two parties Alice and Bob share an entangled state
of the form
\begin{eqnarray}
|\psi'\rangle = \frac{1}{\sqrt{2}}[|00\rangle+|11\rangle]
\end{eqnarray}
If once again one can think of that the asking the question $X$ to
one of the players is equivalent of measuring the spin of the
qubit along x direction (i.e he/she will measure $\sigma_x$), then
the answers to the question are the measurement results +1 and-1.
If $\prod_{i=1}^n\sigma_{x,i}^A\sigma_{x,i}^B$ is the operator to
be applied to the state $|\psi'\rangle$, we obtain,
\begin{eqnarray}
\prod_{i=1}^n\sigma_{x,i}^A \otimes
\sigma_{x,i}^B|\psi'\rangle=|\psi'\rangle
\end{eqnarray}
(where i denotes the number of trials). This clearly indicates
that the operator $\prod_{i=1}^n\sigma_{x,i}^A\sigma_{x,i}^B$ on
its action on the state $|\psi'\rangle$ gives the state back with
eigen value 1, satisfying  the
winning condition in all possible cases.\\
{\bf Game 3:} This is the most important of all the two party non
local games discussed so far. The speciality of the game lies in
its impossibility in the classical settings not only
deterministically but even probabilistically, but still we can
find a quantum strategy by which the game can be won. Quite alike
to the previous cases here we have also considered two players
Alice and Bob who are not allowed to communicate at any stage of
the game. The promise of the game is that if one of the players
are asked to answer the question $\textbf{X}$, the other will be
asked the question $\bar{\textbf{X}}$ which is just the
complementary question to $\textbf{X}$. The solution to both the
questions $\textbf{X}$ and $\bar{\textbf{X}}$ is taken from the
set $G=\{1,-1\}$. The elements $1$ and $-1$ are additive inverse
of each other in the group $(\textbf{\textrm{R}},\textbf{+})$,
where $\textbf{+}$ is the normal arithmetical addition. Without
any loss of generality let us assume that Alice was asked to
answer the question $\textbf{X}$ and Bob was asked to answer the
question $\bar{\textbf{X}}$. Now Alice can choose the answer from
any two elements of the set G (as both of them are feasible
answers to the question $\textbf{X}$). Similarly Bob can also
choose the answer of his question $\bar{\textbf{X}}$ from the same
set G. The players are said to be winners of the game if the
product of their answers is equal to $1$, provided that answer
given by one player is the additive inverse of the answer given by
another player. Mathematically, the winning condition can be
written as,
\begin{eqnarray}
X^A\bar{X}^B=1
\end{eqnarray}
where the output $\bar{X}^B$ of Bob is the additive inverse of the
output $X^A$ given by A. The game is interesting in the sense that
the winning condition is set up in such a way that it is
impossible to achieve in the classical world, even if they are
allowed to communicate. This is because of the fact this is an
impossible event, the situation will never arise when the product
of two additive inverses will be equal to 1.\\
However quantum mechanically, with a prior share of quantum
entanglement this can be easily achieved. Let us consider two
players Alice and Bob are sharing an entangled state of the form,
\begin{eqnarray}
|\psi'\rangle = \frac{1}{\sqrt{2}}[|00\rangle+|11\rangle]
\end{eqnarray}
Now if we consider the asking the question $\textbf{X}$ to the
quantum players is equivalent of saying that they are measuring
the spin of their qubit along x direction, i.e he/she will measure
$\sigma_x$, then in quantum mechanical context the complementary
question $\bar{\textbf{X}}$ will be the measuring same $\sigma_x$.
This is because of the fact $\sigma_x^2=I$(where $I$ is the
identity operator). Measurement result ($+1$ and $-1$ eigen
values)will be the answer. It is clearly evident already from the
previous subsection that.
\begin{eqnarray}
\sigma_{x}^A \otimes \sigma_{x}^B|\psi'\rangle=|\psi'\rangle
\end{eqnarray}
It is evident from the above equation the winning condition is
always achieved, even when the eigen values of the respective
operators are additive inverse of each other. Thus we see that
this particular game attains the winning condition quantum
mechanically, even when it is impossible to do so in the classical
setting .
\section{Conclusion:} In summary we can say that here in this work
we proposed three two party non local games, which doesn't admit
of a winning solution in the classical world at least
deterministically. Interestingly players can win this game if they
share a prior entangled state in between them. These games are
different from the existing pseudo-telepathy games in the sense
that these games do not involve any kind of projective
measurements in the computational basis. Not only that for these
games we have used entangled states of dimension $2\times 2$.
However for pseudo telepathy games this is not possible [10]. Also
for these non local games we not allowing classical players to
communicate at any stage of the game. However in pseudo-telepathy
games classical players are allowed to communicate before the
onset of the game. In the first two games one can win the game not
deterministically, but with a certain probability of success.
Interestingly this probability of success is small even when the
input is very small. However the third game is different from the
first two in the sense that there doesn't exist any classical
setting by which one can win the game, but with a prior share of
entanglement, the quantum players will be successful in winning
the game in all possible situations. One can also implement the
concept of these non local games introduced in this work in multi
party settings.
\section{Acknowledgement:}
 I.C acknowledges Prof C.G.Chakraborti, S. N Bose Professor of
Theoretical Physics, University of Calcutta, for being the source
of inspiration in carrying out research.
\section{Reference:}
$[1]$ G. Brassard, \textit{Foundation of Physics} \textbf{33}(11):
1593-1616, 2003.\\
$[2]$ H.Buhrman, R.Cleve, \textit{Physical Review A} \textbf{56}:
1201-1204, 1997.\\
$[3]$ Vaidman, \textit{Found. Phys.} \textbf{29}: 615 (1999)\\
$[4]$ G.Brassard.etal, Multi-Party Pseudo Telepathy, e-print:
quant-ph/0306042.\\
$[5]$ N.D.Mermin, \textit{American Journal of Physics}
\textbf{58}(8):731-734, 1990.\\
$[6]$ G.Brassard.etal, Quantum Pseudo-Telepathy, e-print:
quant-ph/0407221.\\
$[7]$ Z.Bar-Yossef.etal, \textit{Proceedings of the 36th Annual
ACM Symposium on Theory of Computing}, 128-137, 2004.\\
$[8]$ P.K.Arvind, \textit{Foundations of Physics Letters}
\textbf{15}(4): 397-405, 2002.\\
$[9]$ P.K.Arvind, A simple demonstration of Bell's theorem
involving two observers and no probabilities or inequalities,
e-print: quant-ph/0206070.\\
$[10]$ Bassard.et.al, Minimum entangled state dimension required
for pseudo-telepathy, e-print: quant-ph/0412136.\\
$[11]$ L.Vaidman,\textit{ Phys. Lett.  A} \textbf{286}: 241
(2001).\\
$[12]$ A.Cabello, \textit{Phys. Rev. Lett.} \textbf{86}: 1911
(2001); \textbf{87}: 010403 (2001).\\
$[13]$ A. Cabello, \textit{Phys. Rev. A} \textbf{73}:
022302(2006).

\end{document}